\documentclass[%
reprint,
superscriptaddress,
nofootinbib,
amsmath,amssymb, amsfonts,
aps,
]{revtex4-1}
\usepackage[latin1]{inputenc}
\usepackage{graphicx}
\usepackage{dcolumn}
\usepackage{enumitem} 
\usepackage{bm}
\usepackage{physics}
\usepackage{qcircuit}
\usepackage{amsthm}
\usepackage{mathtools}
\usepackage{nicefrac}
\usepackage{mathrsfs}
\usepackage{algorithm,algpseudocode}
\usepackage{algorithmicx}
\usepackage{float}
\usepackage{subfigure}
\usepackage{booktabs}

\usepackage{hyperref}
\usepackage[noabbrev]{cleveref}

\hypersetup{
	colorlinks,
	linkcolor={blue},
	citecolor={blue},
	urlcolor={blue}
}

\usepackage{tcolorbox}

\newtcolorbox[blend into=figures]{greyfigure}[2][]{float=htb,capture=hbox,
blend before title=colon hang,title={#2},every float=\centering,#1}

\newcommand{\bb}{\mathbb}
\newcommand{\mc}{\mathcal}

\newtheorem*{restate*}{Theorem Restatement}

\newtheorem{example}{Example}

\DeclarePairedDelimiter\ceil{\lceil}{\rceil}

\newcommand{\density}[1]{\left\lvert #1 \right\rangle \!\! \left\langle #1 \right\rvert}
\newcommand{\rega}{\rm A}
\newcommand{\regb}{\rm B}
\newcommand{\regc}{\rm C}
\newcommand{\noise}{\mathcal{E}}
\newcommand{\prepgeneral}{\bar{\sigma}}
\newcommand{\prepstandard}{\bar{\mathcal{U}}}
\newcommand{\preppauli}{\bar{\mathcal{V}}}
\newcommand{\prepcombined}{\bar{\mathcal{W}}}
\newcommand{\repgeneral}{\sigma}
\newcommand{\repstandard}{\mathcal{U}}
\newcommand{\reppauli}{\mathcal{V}}
\newcommand{\repcombined}{\mathcal{W}}
\newcommand{\sprob}[1]{p(#1)}
\definecolor{hpCol}{RGB}{150, 0, 150}

\newcommand{\hphide}[1]{}

\begin{document}
	
	\title{Non-Exponential Behaviour in Logical Randomized Benchmarking}
	
	\author{Athena Ceasura}
	\author{Pavithran Iyer}
	\affiliation{Institute for Quantum Computing, University of Waterloo, ON, N2L 3G1 Canada}
	\author{Joel J. Wallman}
	\affiliation{Keysight Technologies Canada, Kanata, ON K2K 2W5, Canada}
	\author{Hakop Pashayan}
	\affiliation{Institute for Quantum Computing, University of Waterloo, ON, N2L 3G1 Canada}
	\affiliation{Department of Combinatorics and Optimization, University of Waterloo, ON, N2L 3G1 Canada}
    \affiliation{Perimeter Institute for Theoretical Physics, Waterloo, ON, N2L 2Y5 Canada}
    \affiliation{Dahlem Center for Complex Quantum Systems, Freie Universit\"{a}t Berlin, Germany}
	
	\date{\today}
	
	\begin{abstract}
		We construct a \emph{gate and time-independent} noise model that results in the output of a logical randomized benchmarking protocol oscillating rather than decaying exponentially. To illustrate our idea, we first construct an example in standard randomized benchmarking where we assume the existence of ``hidden'' qubits, permitting a choice of representation of the Clifford group that contains multiplicities. We use the multiplicities to, with each gate application, update a hidden memory of the gate history that we use to circumvent theorems which guarantee the output decays exponentially. In our focal setting of logical randomized benchmarking, we show that the presence of machinery associated with the implementation of quantum error correction can facilitate non-exponential decay. Since, in logical randomized benchmarking, the role of the hidden qubits is assigned to the syndrome qubits used in error correction and these are strongly coupled to the logical qubits via a decoder.
	\end{abstract}

	\maketitle

	\section{Introduction} \label{sec:Intro}
	Since the technological development of quantum computers; characterizing noise within a quantum device has become an active and pressing area of study. Randomized Benchmarking (RB) offers an effective and well studied~\cite{Emerson2005, Magesan2011, Magesan2012a, Magesan2012b, Wallman2014a, Wallman2015b, Wallman2016, Wallman2016a, Wallman2018, Helsen2019, Helsen2020AGF} approach to the characterization of noise. RB is a class of protocols that provide a means by which one can compute a figure of merit indicating the `average level of noise' affecting the quantum gates acting at the physical level, i.e. the average fidelity. RB's scalability and robustness to state preparation and measurement (SPAM) errors have made it a ubiquitous noise characterization technique. 
	
	To store and process logical data robustly, quantum devices will likely encode and protect logical data by using some form of quantum error correction (QEC). Logical randomized benchmarking (LRB) aims to characterize the noise afflicting the logical qubits directly. Combes et. al.~\cite{Combes2017} showed that the RB procedure could, with minor modification, be applied at the logical level and still work well assuming simple noise models on both the logical operations and the operations associated with QEC - the measurement of syndromes and the implementation of a decoding operation. However, as we will see, the increased complexity associated with the QEC machinery can bring with it new modes of failure. These new modes of failure show that characterization of noise at the logical level is a more nuanced task that may requiring more robust techniques than those originally designed for characterizing noise at the physical level.
	
	Although there are many variants of RB, typically the procedure involves starting in some initial fiducial state, applying a randomly sampled sequence of $m$ Clifford gates followed by the Clifford gate that is the sequence inverse. The fidelity between the initial and final state is empirically estimated and plotted as a function of the sequence length $m$. Assuming that the action of the Clifford gates on the underlying Hilbert space satisfies certain properties, the output will be a linear combination of exponential decays in the RB sequence length \cite{Wallman2018}. By separating each of these exponential decays in the output data via filter functions, the decay parameters can be determined~\cite{Helsen2020AGF}. In the case where there is a single decay parameter, it will be proportional to the average fidelity.
	
	To analyse the behaviour of RB, the implementation of a Clifford group elements is modelled as a map on the quantum system's Hilbert space that consists of two components; a unitary channel isomorphic to the Clifford group and a noise channel. The first of these is defined by a Clifford group representation. This representation can be decomposed into a sum of irreducible components. In general, the same irreducible component may appear more than once in the sum, if this is not the case, the representation is said to be \emph{multiplicity free}. The second component models the action of noise. In the most general setting, the noise component will depend on the context of the implementation (non-Markovian noise) e.g. the noise may depend on the Clifford group element being implemented (gate dependent noise) or it may depend on when the element is being implemented (time dependent noise). The output of RB is a linear combination of exponential decays provided that the representation is multiplicity free and that the noise is time and gate independent\footnote{Various relaxations of this assumption have also been studied in Refs.~\cite{Wallman2014a, Wallman2018, Helsen2019}.}~\cite{Helsen2020AGF}. 
	
	In the absence of these assumptions, an RB procedure can fail in the sense that the output may not be a linear combination of exponential decays. This failure can be severe, even resulting in output that is an increasing function of sequence length; a phenomenon known as upticks. RB can fail due to the presence of strong non-Markovian noise or, more inconspicuously because the implemented representation contains multiplicity. We use a simple example of the latter in the case of gate and time independent noise to produce periodic oscillations in RB output. Through this example we demonstrate how strong interactions between a device's qubits and the environment can give rise to a memory which facilitates unwanted non-Markovian effects and produces upticks in RB output. In the setting of RB, one may excuse such modes of failure since experimentalists are well aware of and work hard to eliminate strong interactions between the target system (physical qubits of a quantum computer) and the environment. However, in the setting of LRB, the target system is the set of logical qubits and thus is not a physical system that can be isolated from the environment. We show that in the setting of LRB, the additional qubits required for quantum error correction (QEC) can play the role of unwanted environment qubits that are strongly coupled to the target system. Further, the decoder itself can function as a noise process that strongly couples the target system to the environment.
	
    Our work presents a somewhat contrived example that is simple, generalizable and results in the most extreme failure of (L)RB under a very ``RB-friendly'' noise model; gate and time independent noise.

    \section{Notation} \label{sec:notation}
	We use $\mc{P}_n$ to denote the $n$-qubit Pauli matrix group and $\mc{P}_n^+$ to denote the Paulis modulo phases. We denote the \emph{abstract} $n-$qubit Clifford group by $\mc{C}_n$. We use $\mathcal{H}$ to denote the Hilbert space containing all density matrices that may describe the quantum state of the physical system. We denote noise by the map $\noise\in GL(\mathcal{H})$ where $GL(\mathcal{H})$ denotes the space of linear maps from $\mathcal{H}$ to itself. We use $\prepgeneral:\mc{C}_n \rightarrow GL(\mathcal{H})$ to denote a projective group representation of $\mc{C}_n$ acting on $\mathcal{H}$. We use a bar to distinguish the projective representations $\prepgeneral:\mc{C}_n \rightarrow GL(\mathcal{H})$ that act as superoperators from their corresponding matrix representations $\repgeneral:\mc{C}_n \rightarrow \mathcal{H}$. More generally, for any predefined unitary map $U$, we will use $\bar{U}$ to denote the superoperator that acts on a density state $\rho\in \mc{H}$ by conjugation i.e. $\bar{U}\rho:= U\rho U^{\dagger}$.
	
    \section{Randomized Benchmarking} \label{sec:RB}
    Randomized benchmarking (RB) encompasses the most widely studied experimental protocols to estimate the quality of hardware components in a quantum computer. The protocol outlined in~\cref{algo:RB} can be summarized as follows. We prepare a fixed initial state of the system (which, without loss of generality can be) $|0\rangle^{\otimes n}$, apply a sequence of uniformly randomly sampled Clifford gates $G_{1}, \ldots, G_{m-1} \in \mc{C}_n$, and invert the sequence with the last gate 
    $G_m=G_{\mathrm{inv}}:=(G_{m-1}\ldots G_{1})^{-1}$.
    Finally, we measure the state with the POVM $\{I-\density{0}^{\otimes n}, \density{0}^{\otimes n}\}$. In the absence of errors, the $\density{0}^{\otimes n}$ outcome will always be observed. For each sequence length, $m$, taking a statistical average of the outcome over many independent random sequences provides an estimate of the \emph{survival probability} of the input state: $\sprob{m}$, as a function of the sequence length and the average gate infidelity of the underlying faulty gates. The key observation for RB is that there is an exponential decay law relating the length of the sequence of random Clifford gates and the survival probability of the input state, i.e.
    \begin{align}
    \sprob{m} = \sum_{i=1}^r a_i \lambda_i^{m} \label{eq:expoential_decay}
    \end{align}
    for some constants $a_i$ and $\lambda_i$. For each $i$, the quantity of interest, $\lambda_i$, can be estimated using established techniques~\cite{Helsen2020AGF}.

    The theory of RB can be explained in general terms using representation theory. We will outline the key ideas since it paves the road to identifying settings in which the theory breaks down, yielding non-exponential decays, contrary to eq. \ref{eq:expoential_decay}. In general, Clifford gates $G_{1}, \ldots, G_{m}$ applied on a fixed input state $Q$ can be realized using any representation of the Clifford group $\prepgeneral_G$. Using this we write an RB sequence with $m-1$ random Clifford realizations as $\noise \prepgeneral_{G_m} \ldots \noise \prepgeneral_{G_1}$ where we have modelled a faulty implementation as the projective representation $\prepgeneral$ followed by a context independent noise model $\noise$ which does not depend on time, round number, which gate is applied etc..
    
        \begin{algorithm}[H]
        \caption{RB Sequence \label{algo:RB}}
        \begin{algorithmic}[1]
        \algrenewcommand\algorithmicrequire{\textbf{Input:}}
        \algrenewcommand\algorithmicensure{\textbf{Output:}}
        \Require $m\in \mathbb{N}$ and $m > 1$ - sequence length.
        \Ensure An outcome $\hat{o}$.
        \Statex 
        \State Prepare initial state $Q=\ketbra{0}^{\otimes n}$.
        \For{$1 \leq i \leq m - 1$}
        \State    Sample $G_i \in \mc{C}_n$ uniformly.
        \State    Apply the map $\noise \prepgeneral_{G_i}$.\label{step:apply element}
        \EndFor
        \State Set $G_{m}=G^{-1}_1 G^{-1}_2 \ldots G^{-1}_{m-1}$.
        \State Apply the map $\noise\prepgeneral_{G_{m}}$.\label{step:apply inv}
        \State Measure POVM $\{I - Q, Q \}$.
        \State Record outcome as $\hat{o}\in \{0,1\}$ \Comment{$1$ corresponds to $Q$}
        \State \textbf{return} {$\hat{o}$}.
        \end{algorithmic}
        \end{algorithm}
    
    We note that in the case that the RB sequence is executed on a physical device, the representation $\prepgeneral$ and noise map $\noise$ are implicit to the device and unknown. However, a standard choice for $\prepgeneral_G$ is $\prepstandard_G$ which conjugates density states by $\repstandard_G$, the standard matrix representation of $\mc{C}_n$ (see~\cref{sec: reps_details} for a precise definition of these representations).
    
    For each sequence length $m$,~\cref{algo:RB} is repeated many times and the empirical average of the outputs $\hat{o}$ closely approximates the expectation value $\mathbb{E}(\hat{o})$ also known as the survival probability $\sprob{m}$
    \begin{align}
    \sprob{m} &=\frac{1}{|\mc{C}_n|^m}\sum_{G_1,\ldots,G_m} \tr(Q ~\noise\prepgeneral_{G_{\mathrm{inv}}}\prod_{i=1}^m \noise \prepgeneral_{G_i} Q)\notag\\
    &=\tr(Q \noise \mathcal{T}^m Q)\label{eq: exp_val_o}
    \end{align}
    where $Q=\ketbra{0}^{\otimes n}$ and
    \begin{align*}
        \mathcal{T}=\frac{1}{\abs{\mc{C}_n}} \sum_{G\in \mc{C}_n} \prepgeneral_G\noise\, \prepgeneral_G^{-1}
    \end{align*}
    is the \emph{twirl} of the noise map $\noise$. It is easy to show that $\mc{T}$ commutes with $\prepgeneral_G$ for all $G\in \mc{C}_n$. If the decomposition of $\prepgeneral$ into irreducible representations is multiplicity free then by Schur's lemma, there is a decomposition of $\mc{H}$:
    \begin{align}
        \mc{H}=\bigoplus_{i=1}^r \mc{H}_{i}
    \end{align}
    such that $\mc{T}$ acts on $\mc{H}$ by acting on each component $\mc{H}_i$ by scalar multiplication $v\in \mc{H}_i \mapsto \lambda_i v$. 
    Here $r$ is the number of irreducible representations appearing in the decomposition of $\prepgeneral$
    and $\lambda_i$ are real constants determined by $\noise$ and $\prepgeneral$. 
    Substitution into Eq.~\eqref{eq: exp_val_o} shows that as a function of $m$, the expectation value $\mathbb{E}(\hat{o})$ is a sum of $r$ exponential functions with decay rates $\lambda_1, \ldots, \lambda_r$. Each decay rate can be interpreted as the average fidelity rate for each irreducible subrepresentation of the implemented group~\cite{Helsen2020AGF}.

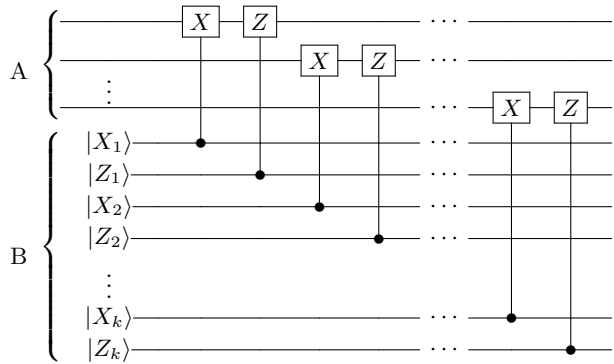
\begin{figure*}
	    \[\begin{array}{ccc}
	       \Qcircuit @C=1em @R=.33em {
            \lstick{} & \qw & \qw & \gate{X} & \gate{Z} & \qw & \qw & \qw & \cdots & & \qw & \qw & \qw\\
            \lstick{} & \qw & \qw & \qw & \qw & \gate{X} & \gate{Z} & \qw & \cdots & & \qw &  \qw & \qw\\
            \lstick{} & \vdots & \\
            \lstick{} & \qw & \qw & \qw & \qw & \qw & \qw & \qw & \cdots & & \gate{X} & \gate{Z} & \qw\inputgroupv{1}{4}{.8em}{2.0em}{\rega}\\
            \lstick{} & \push{\ket{X_1}} & \qw & \ctrl{-4} & \qw & \qw & \qw & \qw & \cdots & &  \qw & \qw & \qw &\rstick{}\\
            \lstick{} & \push{\ket{Z_1}} & \qw & \qw & \ctrl{-5} & \qw & \qw & \qw & \cdots & &  \qw & \qw & \qw\\
            \lstick{} & \push{\ket{X_2}} & \qw & \qw & \qw & \ctrl{-5} & \qw & \qw & \cdots & &  \qw & \qw & \qw\\
            \lstick{} & \push{\ket{Z_2}} & \qw & \qw & \qw & \qw & \ctrl{-6} & \qw & \cdots & & \qw & \qw & \qw\\
            \lstick{} & \push{\vdots} &\\
            \lstick{} & \push{\ket{X_k}} & \qw & \qw & \qw & \qw & \qw & \qw & \cdots & & \ctrl{-6} & \qw & \qw \\
            \lstick{} & \push{\ket{Z_k}} & \qw & \qw & \qw & \qw & \qw & \qw & \cdots & & \qw & \ctrl{-7} & \qw\inputgroupv{5}{11}{.8em}{4.6em}{\regb}\\
        }
	    \end{array}\]
	    \caption{An implementation of $\mc{E}$ on the state $\ketbra{\psi}$, which is split into registers $\rega$ and $\regb$ as in~\cref{exam:RBUpticks}. We have labeled the qubits in register $\regb$ to show how one can encode any $k$ qubit Pauli using the basis states of register $\regb$. In the case of~\cref{exam:RBUpticks}, we use the computational basis states of register $\regb$ to store a destabilizer of the state in register $\rega$, which we call the $\rega$-destabilizer. One can see that $\mc{E}$ fulfils the role outlined in step (2c) of~\cref{exam:RBUpticks}, i.e., applying the $\rega$-destabilizer stored in register $\regb$.}
	    \label{fig:defE}
	\end{figure*}

    In our example below, we make an explicit choice of representation $\prepgeneral$ and error map $\noise$. We show that this model produces periodic upticks despite the fact that the error map $\noise$ is gate and time independent. To bypass the assumptions of RB, our choice of representation necessarily contains multiplicities. We stress that these multiplicities may not be apparent to the experimentalist, who will only observe the resultant upticks.
    
    Our examples will work with two registers labeled `$\rega$' and `$\regb$'. Register $\rega$ is a $k$-qubit system which is the intended target of RB. Register $\regb$ is a $2k$-qubit ancillary system (perhaps unknown to the experimentalist) used as a memory. We will choose $\prepgeneral$ and $\mc{E}$ such that the state of register $\rega$ is always a stabilizer state and thus can be represented by a stabilizer group. Associated with this stabilizer group is a unique destabilizer group~\cite{Aaronson2004} which we call the \emph{$\rega$-destabilizers}. All non-identity elements of the destabilizer group anti-commute with at least one element of the stabilizer group and can be thought of as Paulis that will change/introduce an error to the stabilizer state stored in register $\rega$. We will make heavy use of the $\rega$-destabilizers throughout our examples.
    
    Before we begin the example, let's outline the strategy we will be using to induce periodic upticks. With each gate in the RB sequence we alternate between introducing an $\rega$-destabilizer error to register $\rega$ and using the memory stored in register $\regb$ to correct the error. In stark contrast to an exponential decay, this will result in an oscillating survival probability as a function of sequence length (cf. Fig.~\ref{fig:simple_oscillations}).
    
    \begin{example}
    \label{exam:RBUpticks}
        We begin by making an explicit choice for $\prepgeneral$, which will then suggest a form for our error model $\mc{E}$.
        \begin{align}
        \label{eq:RBNoiseModel1}
            \prepgeneral_G=\prepstandard_G \otimes \preppauli_G
        \end{align}
        where, $\prepstandard_G$ is the standard $k$-qubit unitary representation of $\mc{C}_k$ and acts on register $\rega$ and $\preppauli_G$ is a unitary representation of $\mc{C}_k$ that acts on register $\regb$ which will be defined later.
        We will assume register $\rega$ is initialized in the all zeros computational basis state $\ket{0}^{\otimes n}$ and $\regb$ is initialized in a computational basis state to be specified later. In particular, this is of the form; a stabilizer state on register $\rega$ tensored with a computational basis state on register $\regb$. For all $G\in \mc{C}_k$, the map $\noise \prepgeneral_{G}$ will preserve this form.
        
        We now give our example as a 2 step process:
        \begin{enumerate}
            \item At all times, register $\regb$ is in a computational basis state that is an encoding of an $\rega$-destabilizer $P$. We define this encoding in~\cref{eq:encoding}.
            \item With each element $G\in \mc{C}_k$ that is applied (cf. ~\cref{step:apply element} of ~\cref{algo:RB}), the following maps act on registers $\rega$ and $\regb$:
            \begin{enumerate}
                \item The state in register $\rega$ is updated by $\prepstandard_G$.
                \item Register $\regb$ is updated by the map $\preppauli_G$. This acts to transform the register $\regb$ to a new computational basis state that encodes for the updated $\rega$-destabilizer $P\leftarrow G P G^{-1}$.
                \item The map $\noise$ acts on both registers but has the overall effect of ``reading'' the $k$-qubit Pauli $P$ encoded into the state of register $\regb$ and applying it to the state of register $\rega$ as a unitary map $\ketbra{\psi} \rightarrow P\ketbra{\psi}P^{\dagger}$.
            \end{enumerate}
        \end{enumerate}
        It is easy to verify that the error introduced at each time step due to step 2(c) is undone in the following time step. As a consequence, the survival probability of the initial state oscillates between 0 and 1 for alternate sequence lengths, in other words, causing upticks with a period of two. The remainder of this example is dedicated to specifying the remaining details of how this scheme works. Namely, giving explicit definitions for $B(P)$, $\preppauli_G$, and $\mc{E}$.
        
        First of all, we note that a $k$-qubit Pauli $P$ has the form 
        \begin{align}
            P = i^{x\cdot z}X^{x_1}Z^{z_1}\otimes \ldots \otimes X^{x_k}Z^{z_k} \label{eq:encoding}
        \end{align}
        and thus can be encoded as a length $2k$ bitstring $B(P) = (x_1, z_1, \ldots, x_k, z_k)$ and stored in register $\regb$. We fix an arbitrary initial choice of $\rega$-destabilizer $P_0$ and choose the initial state of register $\regb$ to be $\mu(P_0) = \ketbra{B(P_0)}$. We define $\preppauli_G$ as the superoperator that takes $\mu(P) \mapsto \mu(G P G^{-1})$ (see App.~\ref{sec: reps_details} for further details). This ensures that under the action $\prepgeneral_{G}$, register $\rega$ is updated to a new stabilizer state and the $\rega$-destabilizer stored in register $\regb$ is updated to the corresponding $\rega$-destabilizer.
        
        \begin{figure}[h]
            \centering
            \includegraphics[width=\linewidth]{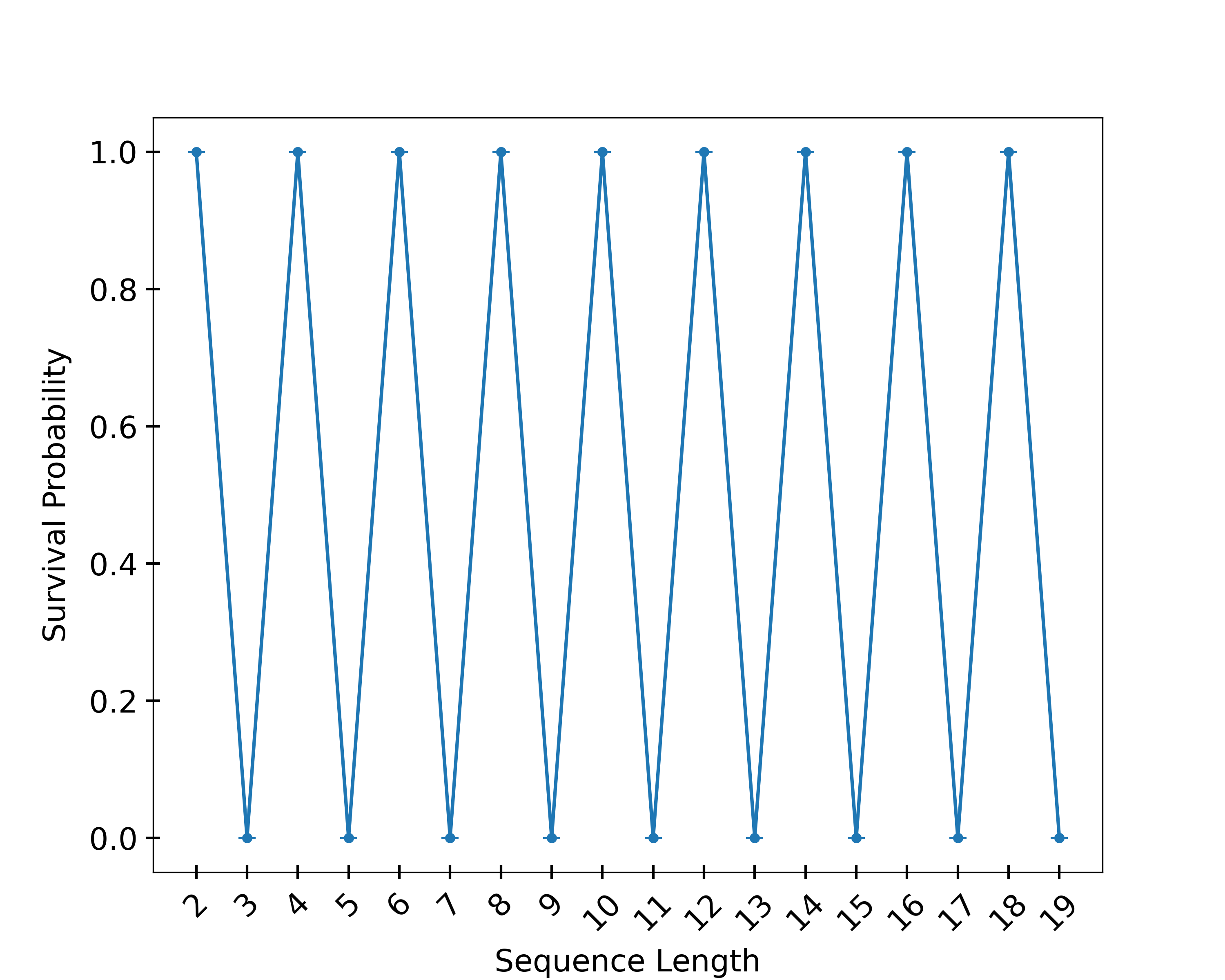}
            \caption{A plot of the survival probability of a simulated RB experiment with the error model presented in~\cref{exam:RBUpticks}. The state is rejected at all odd sequence lengths and accepted for all even sequence lengths. Typically one would observe an exponential decay in the survival probability as sequence length increases, however the non-Markovian effects introduced using register $\regb$ induce oscillations.}
            \label{fig:simple_oscillations}
        \end{figure}
            
        Finally, $\mc{E}$ will induce upticks by enacting the gate stored in register $\regb$ on register $\rega$ each time $\prepgeneral_G$ is applied (cf.~\cref{step:apply element} of~\cref{algo:RB}). We realize $\mc{E}$ using controlled-$X$ and controlled-$Z$ gates which enact respective $X$ or $Z$ gates on register $\rega$, controlled on register $\regb$ see~\cref{fig:defE}. One can verify that $\mc{E}$ is gate and time independent and also fulfills the role prescribed in step (2c) above. In~\cref{algo:RB}, the $\noise$ map is applied a total of $m$ times; $m-1$ times in~\cref{step:apply element} and once in~\cref{step:apply inv}. Thus, the state will be rejected when $m > 1$ is odd because the $\rega$-destabilizer is present in register $\rega$ and accepted when $m$ is even because $\rega$-destabilizer is corrected, yielding upticks in the survival probability.~\Cref{fig:simple_oscillations} shows these oscillations by graphing the survival probability of the state in register $\rega$ as the number of gates in the RB sequence increases.
    \end{example}
    
    We have established that upticks can occur in the above example with a period of two, but we can easily modify our example to exhibit more exotic behaviour. In~\cref{sec:tauperiodupticks} we show that at the cost of additional qubits in register $\regb$, we can delay the application of an $\rega$-destabilizer and produce more complex error patterns.

    \section{Logical Randomized Benchmarking} \label{sec:LRB}
    
    So far, we have only considered the case of RB on physical qubits. But quantum devices will ultimately implement some form of  quantum error correction (QEC), motivating the study of procedures which produce figures of merit for errors at the logical level. Logical Randomized Benchmarking (LRB) is such a procedure introduced in Ref.~\cite{Combes2017} and outlined in~\cref{algo:LRB}. LRB aims to quantify logical error rates by performing a procedure analogous to RB where the randomly sampled Clifford gates are implemented at the logical level; on the code space of a QEC code.

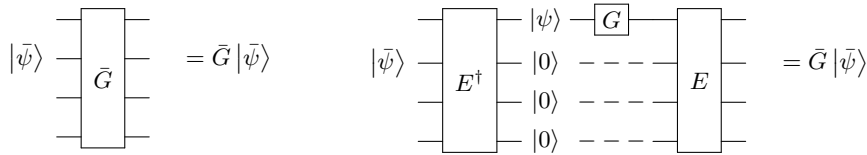
\begin{figure*}
	    \[\begin{array}{ccc}
	       \Qcircuit @C=1em @R=.7em {
          & \multigate{3}{\bar{G}} & \qw &\\
          \lstick{\ket{\bar{\psi}}} & \ghost{\bar{G}} & \qw & \rstick{=\bar{G} \ket{\bar{\psi}}} \\
          & \ghost{\bar{G}} & \qw & \\
          & \ghost{\bar{G}} & \qw & \\
        }
        
        \qquad \qquad \qquad \qquad \qquad
        
	      \Qcircuit @C=1em @R=.7em {
          & \multigate{3}{E^\dagger} & \qw & \ket{\psi} & & \gate{G} & \qw & \multigate{3}{E} & \qw \\
          \lstick{\ket{\bar{\psi}}} & \ghost{E^\dagger} & \qw & \ket{0} &  &  --- & & \ghost{E} & \qw &\rstick{=\bar{G} \ket{ \bar{\psi}}}\\
          & \ghost{E^\dagger} & \qw & \ket{0} & &  --- & & \ghost{E} & \qw \\
          & \ghost{E^\dagger} & \qw & \ket{0} & &  --- & & \ghost{E} & \qw \\
        }
	    \end{array}\]
	    \caption{Copies of the same circuit implementing the logical gate $\bar{G}$. On the RHS, we can see that the encoder $E$ defines a frame that separates the state's syndrome and logical portions. In this ``Unencoded" frame, it is clear that we may, in principle, measure the logical register ($\ket{\psi}$ above) and syndrome register ($\ket*{\vec{0}}$ below) separately. Dashed lines on the RHS remind the reader that the implementation of $\bar{G}$ will be imperfect, so it is likely the noisy implementation of $\bar{G}$ will act to create a non-trivial syndrome.}
	    \label{fig:encodingcircuit}
    \end{figure*}

    QEC works by storing logical quantum information in a subspace (a.k.a. the code space) of the larger physical Hilbert space. For example, an $[[n,k]]$ code stores the logical quantum information, a $k$-qubit quantum state $\ket{\psi}$, in the $n$-qubit physical space by encoding $\ket{\psi}$ as an $n$-qubit quantum state $E (\ket{\psi}\otimes \ket{0}^{\otimes n-k})$ in the code space. The code can be defined by an $n$-qubit unitary called the \emph{encoder} $E$. When $E$ is a Clifford, the QEC code is known as a stabilizer code~\cite{Gottesman1997StabilizerCorrection}.
    The choice of code space aims to ensure that errors acting on the encoded state map the state out of the code space in such a way that this can be detected and any errors introduced to the logical information can be corrected.
    To isolate the action of the physical operations on the logical information, the~\emph{unencoded frame} is particularly useful. In the unencoded frame every physical operation on encoded states is conjugated by the unencoder $E^{\dagger}$, thereby isolating the action of logical operations to the first $k$ qubits, for instance in~\cref{fig:encodingcircuit}. In the unencoded frame, we refer to the first $k$ qubits as the \emph{logical qubits} and the last $n-k$ qubits as the \emph{syndrome qubits}. 
    
    Any operation on the full $n-$qubit space of the form $G \otimes \mc{I}^{\otimes n-k}$ can be realized at the logical level by conjugating by the encoder, $E(G \otimes \mc{I}^{\otimes n-k})E^{\dagger} = \bar{G}$. These operations form ``logical" groups, two of which are relevant here. The first of these is the \textit{Logical Pauli Group} $\bb{L} = \{ E (P \otimes \mc{I}^{\otimes n-k}) E^\dagger~|~\forall P \in \mc{P}_k \}$. 
    The second of these extends $\mathbb{L}$ to a set of operators called the \textit{Logical Clifford Group}, given by $\mc{C}^{\bb{L}} = \{ E (\repstandard_{G} \otimes \mc{I}^{\otimes n-k}) E^\dagger~|~\forall G \in \mc{C}_k \}$.

	When an unknown error $\noise$ affects the encoded state $E (\ket{\psi}\otimes \ket{0}^{\otimes n-k})$, QEC in the stabilizer formalism can be explained as a three step process. First, an error detection step measures the syndrome qubits in the computational basis (in the unencoded frame)\footnote{In practice the syndromes are not measured directly, but the details of this~\cite{Terhal2015QuantumMemories} are unimportant here.} to produce an \emph{error-syndrome} $s~\in~\bb{Z}^{n-k}_{2}$. An outcome $s\neq 0^{n-k}$, detects that the physical state was not in the code space prior to measurement. If this is the case then a logical correction may be required. In the second step, commonly referred to as \emph{decoding}~\cite{Hsieh2011} one computes and applies $D(s)\in \mc{P}_k$ to the logical qubits. Finally, all of the syndrome qubits are reset to $0$, returning the physical state to the code space.~\Cref{algo:QEC} summarizes the key steps in a standard quantum error correction routine.

    \begin{algorithm}[H]
    \caption{Quantum Error Correction (QEC)\label{algo:QEC}}
    \begin{algorithmic}[1]
    \algrenewcommand\algorithmicrequire{\textbf{Input:}}
    \algrenewcommand\algorithmicensure{\textbf{Output:}}
    \Require $D$ - decoder, quantum state.
    \Ensure Restored quantum state.
    \Statex
    \State  Measure the syndrome qubits to obtain syndrome $s$.
    \State  Compute the logical correction $D(s)\in \mc{P}_k$.
    \State  Apply noisy implementation $\noise \prepgeneral (D(s))$.
    \State  Reset the syndrome qubits to $\ket*{\vec{0}}$.\label{step:reset_syndrome}
    \end{algorithmic}
    \end{algorithm}

\begin{algorithm}[H]
\caption{LRB Sequence \label{algo:LRB}}
\begin{algorithmic}[1]
\algrenewcommand\algorithmicrequire{\textbf{Input:}}
\algrenewcommand\algorithmicensure{\textbf{Output:}}
\Require $m\in \mathbb{N}$ and $m > 2$ - sequence length, $\mc{C}$ - Clifford Group, $\prepgeneral$ - representation of $\mc{C}$ acting on all registers, $\noise$ - noise map, $E$ - encoder, $D$ - decoder.
\Ensure An outcome $\hat{o}$.
\Statex
\State Prepare a noisy copy of the ideal initial state \newline $\bar{Q}=E \ketbra{0}^{\otimes n} E^\dagger$.
\For{$1 \leq i \leq m-1$}
\State    Uniformly sample $G_i$ Logical Clifford Group  $\mc{C}$.
\State    Apply noisy implementation $\noise \prepgeneral _{G_i}$.
\State  Apply QEC using $D$. \Comment{see~\cref{algo:QEC}}
\EndFor
\State  Set $G_{m}=G^{-1}_1 G^{-1}_2 \ldots G^{-1}_{m-1}$.
\State  Apply the noisy gate $\noise\prepgeneral_{G_{inv}}$.
\State  Apply QEC with decoder $D$.
\State  Measure POVM $\{ I - \bar{Q}, \bar{Q} \}$ 
\State Record outcome as $\hat{o}\in \{0,1\}$. \Comment{1 corresponds to $\bar{Q}$}
\State \textbf{return} {$\hat{o}$}
\end{algorithmic}
\end{algorithm}
    
Ideally, the representation $\bar{\sigma}$ applied in~\cref{algo:LRB} will be the Logical Clifford Group $\mathcal{C}^{\bb{L}}$. However, in line with~\cref{exam:RBUpticks}, we will now adversarially choose a representation $\prepgeneral$, a time and gate independent noise map $\noise$ and an initial state preperation error to ensure that~\cref{algo:LRB} produces oscillating survival probabilities as a function of sequence length. Like~\cref{exam:RBUpticks}, we will use two registers, $\rega$ and $\regb$; the first to store a stabilizer state and the second to store an $\rega$-destabilizer. However, in the LRB setting, registers $\rega$ and $\regb$ will be defined in the unencoded frame with register $\rega$ corresponding to the logical qubits and register $\regb$ corresponding to the syndrome qubits. This avoids the need for a hidden subspace, but presents two new requirements.

First, we will need to apply the $\rega$-destabilizer stored in register $\regb$ to register $\rega$. In~\cref{exam:RBUpticks}, this was done via the noise map and a similar approach would work here\footnote{This solution is more subtle since we also need to deal with the added complications introduced by the decoder acting on the logical qubits. Without addressing this, the decoder's action will cause the logical state to differ from the target logical state by a Pauli action (the accumulated action of the decoder) resulting in the survival probability not returning to 1. One possible resolution is to require that any Pauli $P\in \mc{P}_k$ applied to the logical qubits by the decoder is accompanied by the update of the $\rega$-destabilzer $Q \leftarrow QP$.}, however we can circumvent the need for this adversarial noise model by taking advantage of QEC.

In particular, by picking a clever encoding for $\rega$-destabilizers in register $\regb$, we employ the decoder to implement the desired logical Pauli. In other words, we can use a decoder specific encoding so that the syndrome measurement outcome $s$ is always such that the correction $D(s)$ chosen by the decoder is the $\rega$-destabilizer. This will require the mild assumption that for each Logical Pauli (modulo phase) $P\in\mc{P}_k^+$ there exists a syndrome $s$ such that $D(s)=P$. We call decoders with this property \emph{surjective}. Using a surjective decoder, one can implement any logical Pauli, in our case an $\rega$-destabilizer, by providing an appropriate syndrome to the decoder $D$. Although the terminology of ``surjective decoder" is new, surjective decoders are very easy to find. For instance, the minimum weight decoder for a CSS code with distance $\geq$ 3 is a surjective decoder, see~\cref{sec:surjectivedecoder} for proof.

Second, we will need to address the syndrome reset since this will erase the gate history information we rely on to generate upticks. For now, we will assume that~\cref{step:reset_syndrome} of~\cref{algo:QEC} completely fails by implementing the identity map on the syndrome qubits instead of resetting them to the all zero state. Then, in~\cref{sec:Circumventing Syndrome Reset}, we will show how this example can be modified to deal with the setting where each qubit is reset with a high reset probability.

\begin{example}
\label{exam:LRBUpticks}
We now demonstrate that under the conditions below, the syndrome measured in LRB can be highly correlated to the implemented gate sequence resulting in oscillating output of an LRB procedure as per~\cref{fig:surivialwithencoding}. We assume the LRB procedure uses an $[[n,k]]$ stabilizer code such that $n > 3k$ and a surjective decoder $D$.
            
Our strategy is inspired by~\cref{exam:RBUpticks}. First of all, let us consider two disjoint subsets of the $n$ physical qubits in the unencoded frame, denoted by registers $\rega$ and $\regb$ respectively. While register $\rega$ contains the $k$ logical qubits, register $\regb$ contains $\geq 2k$ of the syndrome qubits. The size of register $\regb$ is specially chosen to ensure that any logical Pauli operation can be encoded in the computational basis state of a fixed subset of $2k$ qubits in register $\regb$. We construct register $\regb$ out of syndrome qubits, thus, measuring the error syndrome reveals the state of the $2k$ qubits in register $\regb$ to the decoder. 

In~\cref{exam:RBUpticks}, for concreteness, we chose to encode an $\rega$-destabilizer $P$ into register $\regb$ using the state $\ketbra{B(P)}$. Of course, for any $\pi$ that permutes the $2k$ qubit computational basis states, we could have used $\pi \ketbra{B(P)}\pi^{\dagger}$ as the encoding\footnote{With the corresponding adjustment made to the noise map $\noise$.}. We employ this freedom of choice to get the given decoder to apply the $\rega$-destabilizer to the logical qubits. We assume that the decoder satisfies the surjective property. Then the pre-image of any Pauli $P\in \mc{P}_k^+$ may not be unique, however let us choose a particular set of syndromes $S$ which is a possible pre-image of $\mc{P}_k^+$. This defines a unique map $D_{S}^{-1}(P)\in S$. Then there exists a unitary operator $\pi$ (permuting computational basis states in register $\regb$) which acts as $\pi\ket{B(P)} = \ket{D_{S}^{-1}(P)}$ for all $P\in \mc{P}_k^+$. Thus ensuring that a syndrome measurement produces $s=D_{S}^{-1}(P)$ and hence the decoder's correction, $D(s)=P$, is the $\rega$-destabilizer stored in register $\regb$.

Our representation $\prepgeneral:\mc{C}_k\rightarrow (\bb{C}^2)^{\otimes n}$ has the following form:
\begin{align}
\prepgeneral_{G}= \bar{E}^{\dagger} (\prepstandard_{G} \otimes \pi \preppauli_{G} \otimes \bar{\mc{I}}^{\otimes n-3k} \pi^{\dagger}) \bar{E}. \thickspace  \label{eq:gate_error_LRB}
\end{align}
This is equivalent to the representation in~\cref{exam:RBUpticks} (cf.~\cref{eq:RBNoiseModel1}) since they are related by conjugation by a fixed unitary. In this example we will set $\noise \bar{\sigma} = \bar{\sigma}$, as the role of the error model $\noise$ will be taken on by the decoder $D(s)$.

The procedure then works the same as it did for~\cref{exam:RBUpticks}. We assume that the initial state preparation is faulty; producing the initial state 
$E (\ketbra{0}^{\otimes k}\otimes \pi\ketbra{B(P_0)} \pi^{\dagger}) E^\dagger$
where $P_0$ is any destabilizer of $\ket{0}^{\otimes k}$. We note that this assumption can be significantly weakened to realistic probabilistic models of preparation error.
        
While the action of the perfect decoder always resets the error-syndrome, we will for now assume that the syndromes are not reset and address this in the following section. There we will also choose a non-trivial noise map.

The net effect of our choices causes the survival probability to oscillate between 0 and 1, with a period of two.~\Cref{fig:simple_oscillations} is representative of the output of an LRB procedure based on the above assumptions. 
\end{example}
    
\begin{figure}
    \includegraphics[width=\linewidth]{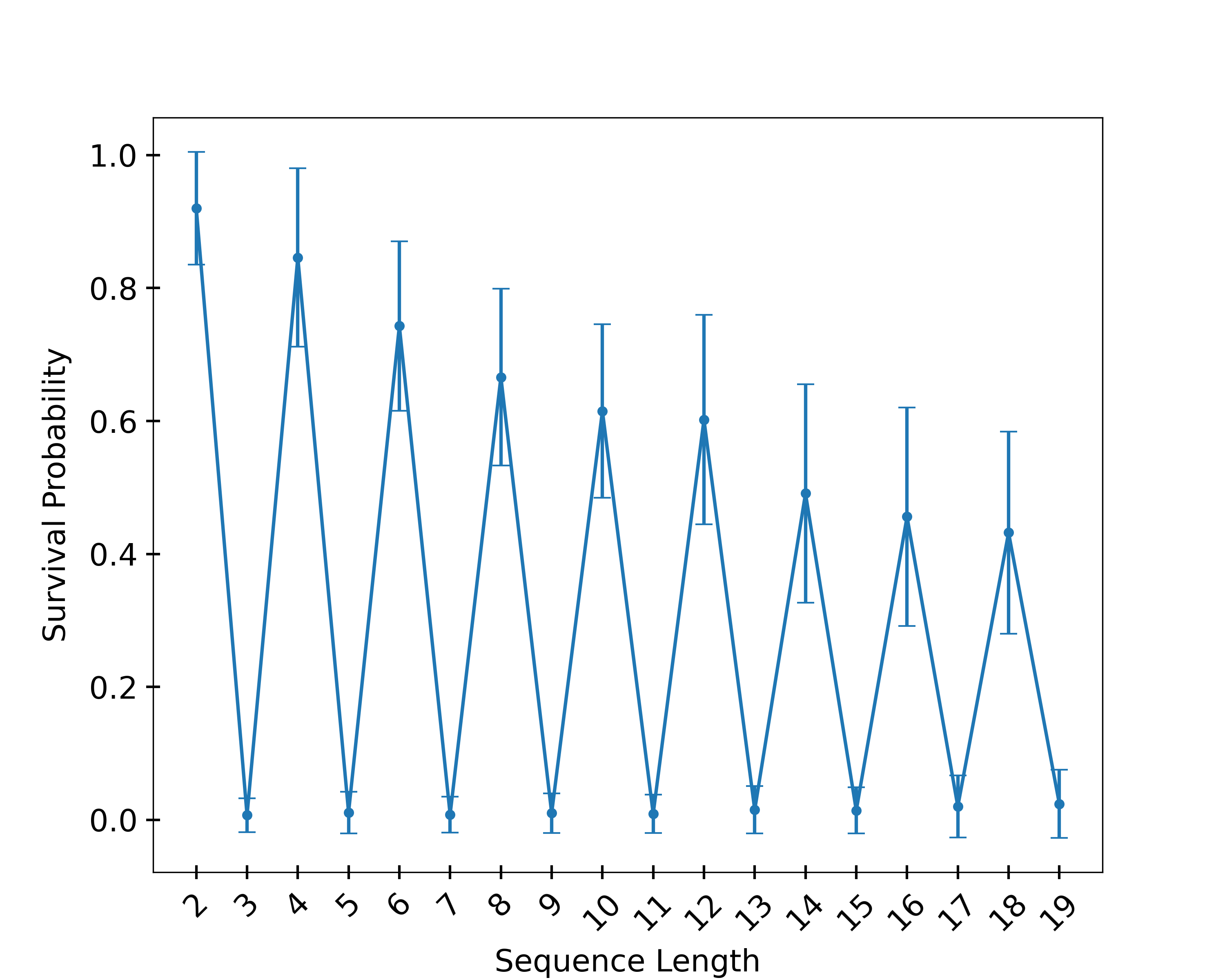}
    \caption{A plot of the survival probability of a simulated LRB procedure with 5 logical qubits and error model given by~\cref{exam:LRBUpticks}. We have introduced a 2\% probability of resetting each of the qubits in register $\regb$ to the $\ket{0}$ state. One can see the amplitude of the oscillations decrease as register $\regb$ probabilistically forgets the binary string which encodes for the $\rega$-destabilizer.}
    \label{fig:lowresetprob}
\end{figure}

\begin{figure}
    \centering
    \includegraphics[width=\linewidth]{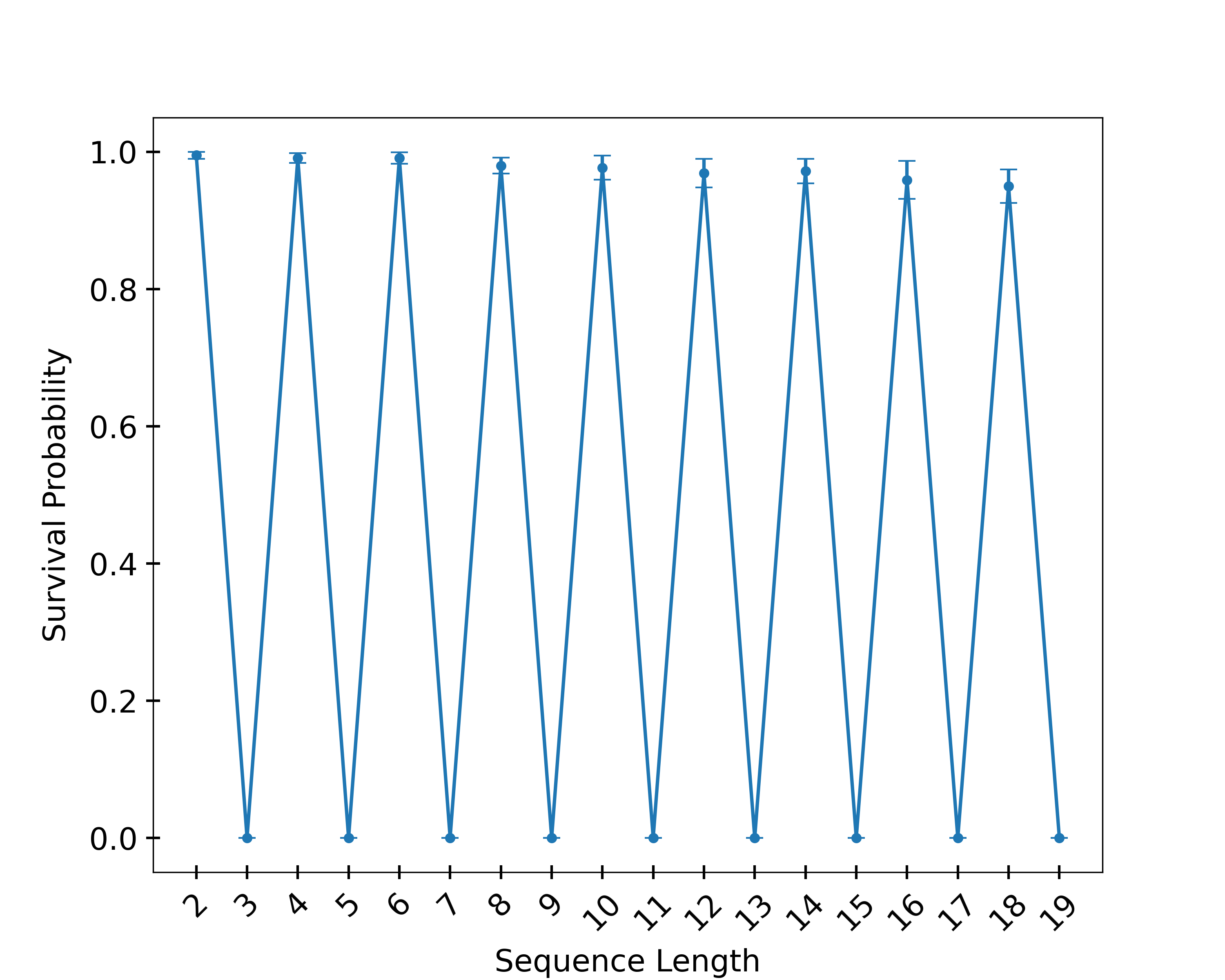}
    \caption{A plot of survival probability in a 20-qubit RB experiment with the noise model given in~\cref{exam:dealwithreset}. In~\cref{exam:dealwithreset}, we used a classical repetition code stored in register $\regb$ of the stabilizer code to circumvent the syndrome reset step of QEC and preserve the classical information which was used to induce non-Markovianity. Each of the qubits in register $\regb$ has a 95\% chance of resetting and we used~\cref{eq:numcopiesapprox} to deduce that 142 copies of register $\regb$ were needed to achieve a 1\% error rate. The persisting oscillations signify that the classical repetition code is preserving the information in register $\regb$, in contrast to~\cref{fig:lowresetprob}.}
    \label{fig:surivialwithencoding}
\end{figure}

~\Cref{fig:lowresetprob} shows this experiment with a small chance of resetting the syndromes. This demonstrates that the oscillating output has some robustness to syndrome reset. We will leverage this fact in the next section to construct a procedure robust to a high probability of syndrome reset.
    
    \section{Circumventing Syndrome Reset} \label{sec:Circumventing Syndrome Reset}
    In the previous section, we used QEC to produce non-Markovian fluctuations in LRB. However, we had to assume that the syndrome reset step of QEC (c.f. \cref{step:reset_syndrome} of \cref{algo:QEC}) entirely fails to occur. In this section, we describe how to circumvent this assumption by embedding a classical error correction code in register $\regb$ to preserve the bitstring which encodes for the $\rega$-destabilizer.
    
        \begin{figure}[h]
        \includegraphics[width=1\linewidth]{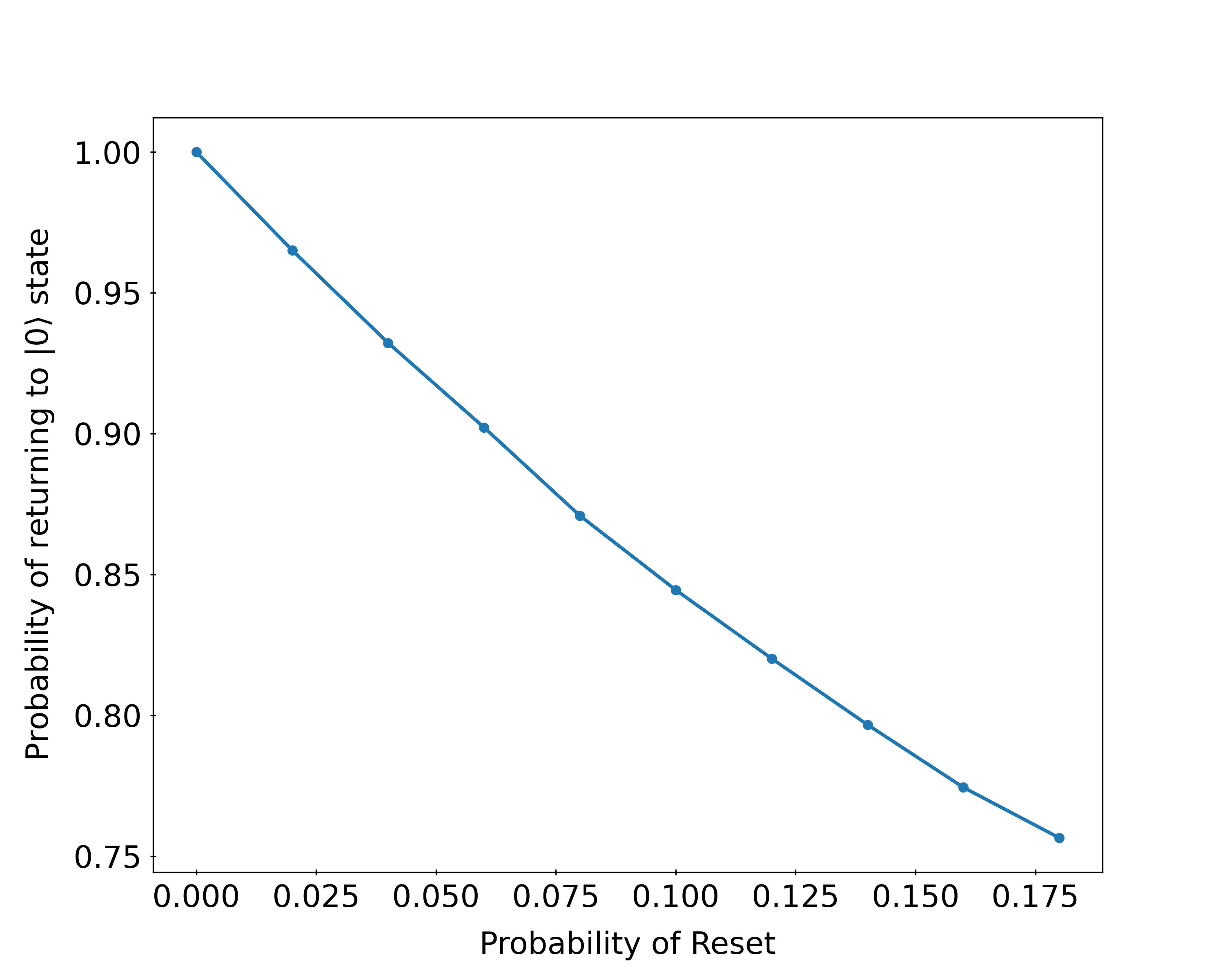}
        \caption{Plots of the survival probability an RB experiment in~\cref{exam:LRBUpticks} after 2 gates were applied. The plot shows the decrease in survival for 5 qubit RB as the probability of resetting each qubit in register $\regb$ increases.}
        \label{fig:returnprob_a}
    \end{figure}
    
    \begin{figure}[h]
        \includegraphics[width=1\linewidth]{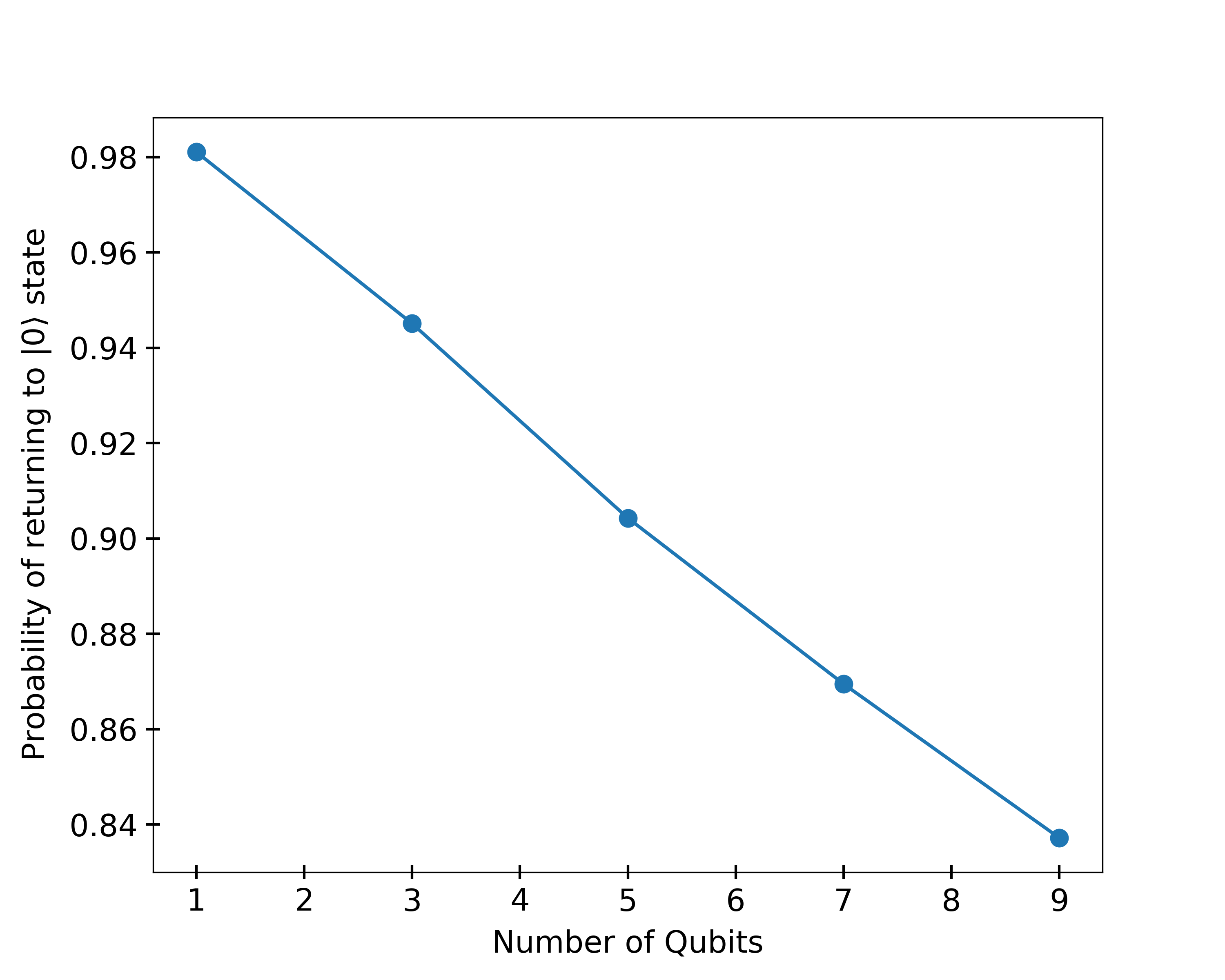}
        \caption{Plots of the survival probability an RB experiment in~\cref{exam:LRBUpticks} after 2 gates were applied. For a constant 2\% reset probability, the plot shows that as the number of qubits increase, a the survival decreases. As register $\rega$ grows, register $\regb$ must grow as well, so one would expect a higher probability that an error occurs on a bit in register $\regb$ such that register $\regb$ no longer encodes for an $\rega$-destabilizer.}
        \label{fig:returnprob_b}
    \end{figure}

    \begin{example}
    \label{exam:dealwithreset}
    We expand upon our previous scheme for generating upticks in LRB given in~\cref{exam:LRBUpticks} where register $\rega$ is given by the logical qubits and register $\regb$ are the syndromes. We begin by specifying the size of the code needed and then go on to construct an explicit noise model which will preserve the $\rega$-destabilizer bitstring stored in register $\regb$ even when QEC resets register $\regb$ with high probability.
    
    We employ an $[[n,k]]$ stabilizer code and a $(2kc, 2k)$ classical repetition code which will be embedded in the syndromes of the $[[n, k]]$ stabilizer code. To accommodate this embedding, we will assume that $n > k(1 + 2c)$ such that the $[[n,k]]$ stabilizer code will have at least $2kc$ syndrome bits. Thus, the syndrome can store $c$ copies of the the $2k$ bit binary string $B(P)$ which encoded for an $\rega$-destabilizer in~\cref{exam:LRBUpticks}. Let $B'_{i, j}(P)$ denote the $j^{th}$ bit in the $i^{th}$ copy of $B(P)$ such that one can view register $\regb$ as containing a binary matrix with a copy of $B(P)$ in each row. 
    
    The representation of the Clifford group for this example is exactly the same as in~\cref{eq:gate_error_LRB}. However the action of $\mc{E}$ is to go iterate each copy in register $\regb$ and see if there is a $1$ in the $j^{th}$ position. If there is a $1$ in $j^{th}$ position of any of the copies, then $\mc{E}$ sets the $j^{th}$ bit in each copy to 1. One can show this error model preserves a bit in $B(P)$ so long as the bit is preserved in a single copy of $B(P)$. Also note this noise model is gate and time independent, meeting the assumptions of RB as in our previous examples.
    
    Let $A$ be the probability that the $\rega$-destabilizer bitstring is preserved after $m$ gates (i.e. roughly the amplitude of the oscillations after $m$ gates). In~\cref{sec:smalldeltaapprox} we show that if $A \geq \sqrt{2} - 1$ and each bit in register $\regb$ is reset independently with probability $r$, we have that the number of copies $c$ required to achieve $A$ is bounded by
    \begin{align}
        c_{k, m}(A) \leq \log_{1/r}\left(\frac{\sqrt{2km}}{1 - A}\right). \label{eq:numcopiesapprox}
    \end{align}
    As the number of physical qubits scales logarithmically, our example is feasible even for very compact codes which require $O(k \log (k))$ physical qubits.
    \end{example}
    
    Thus, we have used a classical repetition code to circumvent syndrome reset and still have oscillations.~\Cref{fig:returnprob_a,fig:returnprob_b} show how the amplitude $A$ decays for small $r$ and $k$, demonstrating that in these regimes oscillations can be preserved using a reasonable number of physical qubits.
    
	One can find the code used for the simulations at~\cite{our_code} where we made use of the code found in~\cite{gf2elim} as a subroutine.

	\section{Conclusion}
	We presented toy scenarios which utilize classical memory to produce large oscillations in the survival probability of RB and LRB experiments. In the setting of RB, this provided an explicit mechanism though which, multiplicities in the representation combined with gate and time independent noise to give rise to non-Markovian effects. In the setting of LRB, our work showed that the presence of additional QEC machinery provides additional mechanisms for the failure of LRB. In particular, the additional qubits associate with QEC can carry a memory that facilitate non-Markovian noise and the decoder itself can function as a noise process that strongly couples the target system to the environment.
	
	Although our examples are somewhat contrived, they result in the most extreme failure of RB. We suspect that there exist realistic noise models which result in less severe failure of RB. Assessing these noise models and providing a more robust theory of non-Markovianity in logical randomized benchmarking remain promising future areas of research.

\section*{Acknowledgements}
We thank Stefanie Beale, Arnaud Carignan-Dugas and Joseph Emerson for useful feedback and discussions. 
Research at Perimeter Institute is supported in part by the Government of Canada through the Department of Innovation, Science and Economic Development Canada and by the Province of Ontario through the Ministry of Colleges and Universities. HP acknowledges the support of the Natural Sciences and Engineering Research Council of Canada (NSERC) discovery grants [RGPIN-2019-04198] and [RGPIN-2018-05188], the BMBF (Hybrid, MuniQC-Atoms, DAQC), the MATH+ Cluster of Excellence, the Einstein Foundation (Einstein Unit on Quantum Devices), the QuantERA (HQCC), the Munich Quantum Valley (K8), and the DFG (CRC 183, EI 519 20-1, EI 519/21-1). HP also acknowledges the Centre for Quantum Software and Information at the University of Technology Sydney for hosting him as a visiting scholar.


	\newpage
	\appendix

    \vspace{3mm}
    
	\newpage
	\section{Definitions of relevant representations}
	\label{sec: reps_details}
	
	In this section we define the two projective representations $\prepstandard$ and $\preppauli$ and their associated representations $\repstandard$ and $\reppauli$ respectively. 
	
	Recall that the $n-$qubit Clifford group $\mathcal{C}_n$ can be generated by the two qubit controlled-NOT gate ($\mathrm{CX}$ gate) and the single qubit Phase and Hadamard gates ($\mathrm{S}$ and $\mathrm{H}$ gates respectively). For $n\in \mathbb{N}$, we define the representation $\repstandard: \mathcal{C}_n \rightarrow GL((\mathbb{C}^2)^n)$ through the standard matrix representations of the generating elements ($\mathrm{CX, S, H}$) of $\mathcal{C}_n$:
	\begin{align*}
	\mathcal{U_{\mathrm{CX}}}=\begin{bmatrix}
    1 & 0 & 0 & 0\\
    0 & 1 & 0 & 0\\
    0 & 0 & 0 & 1\\
    0 & 0 & 1 & 0
    \end{bmatrix}, \quad \mathcal{U_{\mathrm{S}}}=\begin{bmatrix}
    1 & 0 \\
    0 & i 
    \end{bmatrix}, \quad \mathcal{U_{\mathrm{H}}}=\frac{1}{\sqrt{2}}\begin{bmatrix}
    1 & 1 \\
    1 & -1
    \end{bmatrix}.
	\end{align*}
	
	For $p,q\in \mathbb{Z}_2$, we define the single qubit Hermitian operators\footnote{Knowns as the Weyl-Heisenberg displacement operators.}:
	\begin{align}
	    D_{p,q}:=i^{pq}X^p Z^q,
	\end{align}
	where $X$ and $Z$ are the single qubit Pauli $X$ and $Z$ operators. For $p,q\in \mathbb{Z}^n_2$, we define the $n-$qubit counterparts:
	\begin{align}
	    D_{(p,q)}:=\bigotimes_{i\in [n]}D_{(p_i,q_i)}.
	\end{align}
	We now define the map $\reppauli:\mathcal{C}_n\rightarrow GL((\mathbb{C}^2)^{2n})$ as:
	\begin{align}\label{eq:reppauli}
	    \reppauli_G:=\sum_{a,b\in \mathbb{Z}_2^{2n}} 2^{-n}\tr(D_b \, \repstandard_G D_a \, \repstandard_G^{-1})\ket{b}\!\!\bra{a}.
	\end{align}
	We note that the set $\mathbb{P}_n:=\{D_a~|~a\in \mathbb{Z}_2^{2n}\}$ is just the set of $n-$fold tensor products of single qubit identity and Pauli operators $I,X,Y,Z$. Thus, the normalized set $2^{-n/2}\mathbb{P}_n$ forms an orthonormal basis of $GL((\mathbb{C}^2)^{2n})$ with respect to the  Hilbert-Schmidt inner product i.e. $2^{-n}\tr(D_a^{\dagger} D_b)=\delta_{a,b}$. Further, for all $G\in \mathcal{C}_n$ and $D\in \mathbb{P}_n$, up to a sign, the operator $\repstandard_G D\, \repstandard_G^{-1}$ is in $\mathbb{P}_n$. Hence, in the computational basis, $\reppauli_G$ acts as a signed permutation. One can show that the map $\reppauli:\mathcal{C}_n\rightarrow GL((\mathbb{C}^2)^{2n})$ defined in eq.~\eqref{eq:reppauli} is in fact a unitary representation of $\mathcal{C}_n$.
	We define the representation $\repcombined:\mathcal{C}_n \rightarrow GL((\mathbb{C}^2)^{3n})$ using $\repcombined_G:=\repstandard_G \otimes \reppauli_G$. Finally, we extend the representations:
	\begin{align*}
	    \repstandard:\mathcal{C}_n \rightarrow &GL((\mathbb{C}^2)^{n})\\
	    \reppauli:\mathcal{C}_n \rightarrow &GL((\mathbb{C}^2)^{2n})\\
	    \repcombined:\mathcal{C}_n \rightarrow &GL((\mathbb{C}^2)^{3n})
	\end{align*}
	to their corresponding projective representations:
	\begin{align*}
	    \prepstandard:\mathcal{C}_n \rightarrow &GL(GL((\mathbb{C}^2)^{n})); \quad \prepstandard_G (\rho):= \repstandard_G \rho\, \repstandard_G^{-1}\\
	    \preppauli:\mathcal{C}_n \rightarrow &GL(GL((\mathbb{C}^2)^{2n})); \quad \preppauli_G (\rho):= \reppauli_G \rho\, \reppauli_G^{-1}\\
	    \prepcombined:\mathcal{C}_n \rightarrow &GL(GL((\mathbb{C}^2)^{3n})); \quad \prepcombined_G (\rho):= \repcombined_G \rho\, \repcombined_G^{-1}\text{.}
	\end{align*}

	\section{Upticks with Arbitrary Period} \label{sec:tauperiodupticks}
	
	We will show that the upticks in all the examples in this paper can occur with arbitrary period $\repgeneral$ using another register `$\regc$.' Register $\regc$ consists of a $\tau$ level system and a qubit. 
	
	For~\cref{exam:RBUpticks} we will augment~\cref{eq:RBNoiseModel1} as:
    \begin{align}
    \label{eq:RBNoiseModel2}
        \noise \prepgeneral_{G}=\mc{E} (\prepstandard_G \otimes \preppauli_G \otimes \mathfrak{C}),
    \end{align}
    where
    \begin{align*}
        \mathfrak{C} = \sum_{i=0}^{\tau-3} \ketbra{i+1}{i} \otimes \mc{I} + \ketbra{\tau-1}{\tau-2} \otimes X + \ketbra{0}{\tau-1}  \otimes X
    \end{align*}
    is the error model on register $\regc$. Therefore, the $\tau$ level system stores the timestep mod $\tau$. The extra qubit, known as the trigger qubit, will be $1$ after $\tau$ gates are applied and then is immediately flipped back to $0$. It is important to note that the $\tau$ level system could also be created using $\ceil{\log_2{\tau}}$ qubits and a similar map to $\mathfrak{C}$.
    
    To make~\cref{exam:RBUpticks} have period $\tau$ we promote the controlled gates in $\mc{E}$ (see~\cref{fig:defE}) to Toffoli gates, with the new control on the trigger qubit. Now, the $\rega$-destabilizer is only applied when the trigger qubit is 1, that is, every $\tau$ steps, rather than at each step. This leads to the same action as in~\cref{exam:RBUpticks}, except with period $\tau$.

    \section{An example of surjective decoders}
    \label{sec:surjectivedecoder}
	
	Here we prove that the minimum weight decoder is surjective for any non-degenerate [[$n$, 1, $d$]] CSS code with distance at least $3$. Recall that a decoder $D$ is surjective when $\forall L \in \bb{L}$ there exists a syndrome $s_L$ such that $D(s_L) = L$. In order to prove this we will make use of two assumptions about the code:
	\begin{enumerate}
	    \item Logical operations always take the form $P^{\otimes n}$ for $P \in \mc{P}_1$. That is, the code has transversal logical operations.
	    \item We assume that the Pauli errors supported on a single qubit $X_0$, $Y_0$, and $Z_0$, have unique syndromes. Write these syndromes as $s_x$, $s_y$, and $s_z$ respectively. Any non-degenerate code satisfies this assumption.
	\end{enumerate}
	This proof will show that $D(s_x) = X_L$ and similar proofs will follow for $s_y$, and $s_z$. This is sufficient to prove that $D$ is surjective as $s_x$, $s_y$, and $s_z$ collectively allow us to use $D$ to apply any logical operation. Note that the above criterion is sufficient, but not necessary for $D$ to be a surjective decoder.
	
	Since $\bb{L}$ emulates the Paulis on a single qubit, each non-identity element of $\bb{L}$ commutes with itself and anti-commutes with all other non-identity elements of $\bb{L}$. Thus, because $D(s) \in \bb{L}$, we can determine which element of $\bb{L}$ it is based on how it commutes and anti-commutes with the other elements of $\bb{L}$. This principle can be extended to any Pauli, not just with logical Paulis as we have just discussed. In other words, we can determine the effect of any Pauli on the logical qubit by examining how it commutes with the Logical Paulis.
	
	To show this we use the fact every Pauli $P$ has a decomposition into $P = L_{P} P_{rest}$ where $L_{P}\in \bb{L}$ affects the logical qubit and $P_{rest}$ does not affect the logical qubit (i.e. $P_{rest}$ commutes with $\bb{L}$)~\cite{Poulin2006}. Since, $P_{rest}$ commutes with all logical operations we know that $\forall L \in \bb{L}$ $[P, L] = 0 \implies [L_p, L] = 0$ and $\{P, L\} = 0 \implies \{L_p, L\} = 0$. So to determine $L_P$ it is sufficient to check the commutation relations of $P$ with all $L \in \bb{L}$. In other words, to correct the logical error produced by $P$, we should have $D(s_P) = L_P$ where $s_P$ is the syndrome corresponding to $P$. Moreover, we can determine $L_P$ by examining the commutation relations of $P$ with the logical operations as we did in the previous paragraph.
	
	We can now proceed with the proof. As $D(s_x)$ is the minimum weight decoder in a code of distance $\geq$ 3, it must correct $X_0$. To do this, we set $D(s_x)$ to invert the action of $X_0$ on the logical qubit. As suggested above, we can calculate the affect of $X_0$ on the logical qubit by examining the commutation relations of $X_0$ with the logical operations. Using bullet point 1 above, is fairly easy to see that $X_0$ commutes with $X_L$ and anti-commutes with the other two non-identity Logical Paulis. So the logical correction $D(s_x)$ corresponding to $X_0$ must be $X_L$. A similar proof can show $D(s_y) = Y_L$ and $D(s_z) = Z_L$. Since $D$ can implement every logical operation using the syndromes $s_x$, $s_y$, and $s_z$ we have shown $D$ is surjective.
    
	\section{Estimating Number of Copies Needed to Prevent Reset} \label{sec:smalldeltaapprox}
	We now show how to select $c$ such that the $(3kc, 2k)$ classical repetition code in register $\regb$ of~\cref{exam:dealwithreset} is robust to the reset performed by the decoding. If the $j^{th}$ bit is supposed to be $1$ we can write the probability that at least one copy of the $j^{th}$ bit is not reset to $0$ as $\delta_j = 1 - r^c$. Not all bits in $B_i(P)$ should be $1$, but we can use this worst case to bound $A$ with the equation $A \geq (\prod_{j = 1}^{2k}\delta_j)^m$.
	
	Next, if we assume $\delta_j = \delta_k~\forall j,k < 2k$ we have that $A \geq (1 - r^c)^{2km}$. So solving for $c$ gives
	
	\begin{align}
	    \frac{\log_2(1- A^{\frac{1}{2km}})}{\log_2{r}} \geq c \label{eq:easycbound}
	\end{align}
	where the last line uses the fact that $\log_2{r} < 0$ to flip the direction of the inequality. Thus we have found an upper bound for $c$, however~\cref{eq:easycbound} does not show the scaling of $k$ and $m$ clearly. So we will continue on to find the more parseble inequality
	\begin{align}
	    \log_2\left(1 - A^{\frac{1}{2km}}\right) \leq \log_2\left(\frac{1 - A}{\sqrt{2 km}}\right).\label{eq:smalldeltaapprox}
	\end{align}
	First, we note the relation 
	\begin{align}
	    (1 - A) &= (1 - A^{\frac{1}{2}})(1 + A^{\frac{1}{2}})\\
	    &= (1 - A^{\frac{1}{4}})(1 + A^{\frac{1}{4}})(1 + A^{\frac{1}{2}})\\
	    &= (1 - A^{2^{-\ell}}) \prod_{i = 1}^{\ell} (1 + A^{2^{-i}}).
	\end{align}
	Taking the logarithm gives
	\begin{align}
	    \log_2(1 - A) 
	    = \log_2(1 - A^{2^{-\ell}}) + \sum_{i = 1}^{\ell} \log_2(1 + A^{2^{-i}}). \label{eq:smalldeltalemmaexact}
	\end{align}
	Next, we use our assumption that $\sqrt{2} - 1 \leq A < 1$.
	\begin{gather}
	    \sqrt{2} - 1 \leq A^{2^{-i}}\\
	    \sqrt{2} \leq 1 + A^{2^{-i}}\\
	    \frac{1}{2} \leq \log_2(1 + A^{2^{-i}}).
	\end{gather}
	Applying this inequality to~\cref{eq:smalldeltalemmaexact} and rearranging gives
	\begin{gather}
	    \log_2(1 - A) \geq \log_2(1 - A^{2^{-\ell}}) + \frac{\ell}{2} \\
	    \log_2(1 - A^{2^{-\ell}}) \leq \log_2(1 - A) - \frac{\ell}{2}.
	\end{gather}
	Then setting $2^{\ell} = 2km$ gives
	\begin{gather}
	    \log_2(1 - A^{{\frac{1}{2km}}}) \leq \log_2(1 - A) - \frac{1}{2}\log_2(2km).
	\end{gather}
	Which can easily be rearranged into~\cref{eq:smalldeltaapprox}.
	
	Using~\cref{eq:smalldeltaapprox} and~\cref{eq:easycbound} gives
	\begin{gather}
	    c \leq \log_{r}\left(\frac{1 - A}{\sqrt{2km}}\right).
	\end{gather}
	However this form remains unintuitive, as the base of the logarithm is less than 1. A logarithmic identity gives
	\begin{gather}
	    c \leq \log_{1/r}\left(\frac{\sqrt{2km}}{1 - A}\right).
	\end{gather}
	

\begin{thebibliography}{10}

\bibitem{Emerson2005}
J.~Emerson, R.~Alicki, and K.~\.Zyczkowski, ``Scalable noise estimation with
  random unitary operators,'' {\em
  \href{http://dx.doi.org/10.1088/1464-4266/7/10/021}{Journal of Optics B:
  Quantum and Semiclassical Optics}}, {\bfseries 7}, pp.~S347--S352,  (2005).

\bibitem{Magesan2011}
E.~Magesan, J.~M. Gambetta, and J.~Emerson, ``{Scalable and robust randomized
  benchmarking of quantum processes},'' {\em
  \href{http://dx.doi.org/10.1103/PhysRevLett.106.180504}{Physical Review
  Letters}},  (2011).

\bibitem{Magesan2012a}
E.~Magesan, J.~M. Gambetta, B.~R. Johnson, C.~A. Ryan, J.~M. Chow, S.~T.
  Merkel, M.~P. Da~Silva, G.~A. Keefe, M.~B. Rothwell, T.~A. Ohki, M.~B.
  Ketchen, and M.~Steffen, ``{Efficient measurement of quantum gate error by
  interleaved randomized benchmarking},'' {\em
  \href{http://dx.doi.org/10.1103/PhysRevLett.109.080505}{Physical Review
  Letters}},  (2012).

\bibitem{Magesan2012b}
E.~Magesan, J.~M. Gambetta, and J.~Emerson, ``{Characterizing quantum gates via
  randomized benchmarking},'' {\em
  \href{http://dx.doi.org/10.1103/PhysRevA.85.042311}{Physical Review A -
  Atomic, Molecular, and Optical Physics}}, {\bfseries 85}, 4,  (2012).

\bibitem{Wallman2014a}
J.~J. Wallman and S.~T. Flammia, ``{Randomized benchmarking with confidence},''
  {\em \href{http://dx.doi.org/10.1088/1367-2630/16/10/103032}{New Journal of
  Physics}},  (2014).

\bibitem{Wallman2015b}
J.~Wallman, C.~Granade, R.~Harper, and S.~T. Flammia, ``{Estimating the
  coherence of noise},'' {\em
  \href{http://dx.doi.org/10.1088/1367-2630/17/11/113020}{New Journal of
  Physics}},  (2015).

\bibitem{Wallman2016}
J.~J. Wallman and J.~Emerson, ``{Noise tailoring for scalable quantum
  computation via randomized compiling},'' {\em
  \href{http://dx.doi.org/10.1103/PhysRevA.94.052325}{Physical Review A}},
  (2016).

\bibitem{Wallman2016a}
J.~J. Wallman, M.~Barnhill, and J.~Emerson, ``{Robust characterization of
  leakage errors},'' {\em
  \href{http://dx.doi.org/10.1088/1367-2630/18/4/043021}{New Journal of
  Physics}},  (2016).

\bibitem{Wallman2018}
J.~J. Wallman, ``{Randomized benchmarking with gate-dependent noise},'' {\em
  \href{http://dx.doi.org/10.22331/q-2018-01-29-47}{Quantum}},  (2018).

\bibitem{Helsen2019}
J.~Helsen, X.~Xue, L.~M. Vandersypen, and S.~Wehner, ``{A new class of
  efficient randomized benchmarking protocols},'' {\em
  \href{http://dx.doi.org/10.1038/s41534-019-0182-7}{npj Quantum Information}},
   (2019).

\bibitem{Helsen2020AGF}
J.~Helsen, I.~Roth, E.~Onorati, A.~Werner, and J.~Eisert, ``General framework
  for randomized benchmarking,'' {\em
  \href{http://dx.doi.org/10.1103/prxquantum.3.020357}{PRX Quantum}},
  {\bfseries 3},  (2022).

\bibitem{Combes2017}
J.~Combes, C.~Granade, C.~Ferrie, and S.~T. Flammia,
  \href{https://doi.org/10.48550/arXiv.1702.03688}{``Logical randomized
  benchmarking,''}  (2017),
  \href{http://arxiv.org/abs/1702.03688}{arXiv:1702.03688}.

\bibitem{Aaronson2004}
S.~Aaronson and D.~Gottesman, ``Improved simulation of stabilizer circuits,''
  {\em \href{http://dx.doi.org/10.1103/PhysRevA.70.052328}{Physical Review A}},
  {\bfseries 70}, p.~052328,  (2004).

\bibitem{Gottesman1997StabilizerCorrection}
D.~Gottesman,
  \href{https://doi.org/10.48550/arXiv.quant-ph/9705052}{``{Stabilizer Codes
  and Quantum Error Correction},''}  (1997).

\bibitem{Terhal2015QuantumMemories}
B.~M. Terhal, ``Quantum error correction for quantum memories,'' {\em
  \href{http://dx.doi.org/10.1103/RevModPhys.87.307}{Reviews of Modern
  Physics}}, {\bfseries 87}, pp.~307--346,  (2015).

\bibitem{Hsieh2011}
M.~H. Hsieh and F.~Le~Gall, ``{NP-hardness of decoding quantum error-correction
  codes},'' {\em \href{http://dx.doi.org/10.1103/PhysRevA.83.052331}{Physical
  Review A - Atomic, Molecular, and Optical Physics}},  (2011).

\bibitem{our_code}
A.~Ceasura, \href{https://github.com/AthenaCaesura/Non-exponentialRB}{``Code
  used for numerics,''}  (2022).

\bibitem{gf2elim}
S.~Cornell, \href{https://gist.github.com/popcornell}{``popcornell (1.0),''}
  (2018).

\bibitem{Poulin2006}
D.~Poulin, ``{Optimal and efficient decoding of concatenated quantum block
  codes},'' {\em \href{http://dx.doi.org/10.1103/PhysRevA.74.052333}{Physical
  Review A - Atomic, Molecular, and Optical Physics}},  (2006).

\end{thebibliography}
\end{document}